\documentclass[twocolumn,prb,superscriptaddress,nofootinbib]{revtex4-1}
\usepackage[english]{babel}
\usepackage[ansinew]{inputenc}
\usepackage{boldline,multirow,xcolor,colortbl}
\usepackage{times}
\usepackage{graphicx}
\usepackage{graphics}
\usepackage{amsmath}
\usepackage{amsfonts}
\usepackage{amssymb}
\usepackage{amsbsy}
\usepackage{dsfont}
\usepackage{epstopdf}
\usepackage{makeidx}
\usepackage{subfigure}
\usepackage{color} 
\usepackage{pgf}
\usepackage{bm}
\usepackage{tikz} 
\usepackage[normalem]{ulem}
\usepackage[symbol]{footmisc}
\newcommand{\be}{\begin{equation}}
\newcommand{\ee}{\end{equation}}
\newcommand{\bea}{\begin{eqnarray}}
\newcommand{\eea}{\end{eqnarray}}
\newcommand{\blue}[0]{\color{black}{\rm}}

\setcitestyle{open={[},close={]}}
\begin{document}

\title{Nonadiabatic dissociation of molecular Bose-Einstein condensates: competition between chemical reactions}

%
\author{Rajesh K. Malla}
\affiliation{Center for Nonlinear Studies and Theoretical Division, Los Alamos National Laboratory, Los Alamos, New Mexico 87545, USA}
\begin{abstract}
We provide a framework to solve generic models describing the dissociation of multiple molecular Bose-Einstein condensates in a nonadiabatic regime. The competition between individual  chemical reactions can lead to non-trivial dependence on critical components such as path interference and symmetries, thus, affecting the final  distribution of atomic population. We find an analytical solution for an illustrative example model involving four atomic modes. When the system parameters satisfy $CPT$ symmetry, where $C$ is charge conjugation, $P$ is parity, and $T$ is time-reversal symmetry, our solution predicts a population imbalance between atomic modes that is exponentially sensitive to system parameters. However, a weakly broken symmetry alters the population in each atomic mode and can reverse the population imbalance. Our solution also demonstrates a strong quantum correlation between atomic modes that leads to the spontaneous production of atoms in a multi-mode squeezed state. Moreover, in our framework, a time-dependent non-Hermitian quantum mechanics naturally manifests which can alternatively be realized experimentally in photonic systems.
\end{abstract}
\maketitle
\section{Introduction}
Recent experimental advances in ultracold atomic and molecular gases enable us to investigate many-body precision physics, with potential application in quantum control of chemical reactions, precision measurements, quantum simulation and quantum information processing\cite{reviewcold,Carr_2009,coldcontrol}. Ultracold platforms allow control of interaction parameters that has lead to some fascinating experimental observations including, reaction between atomic and molecular Bose Einstein condensates (BEC) ~\cite{exp-21nat}, realization of Unruh radiation~\cite{Unruh}, and Bose Fireworks\cite{Bosefireworks}.  Over the past two decades, there has been a tremendous surge in the theoretical ~\cite{yurovsky,altland-LZ,DTCM1,itin,Malla2021} and experimental ~\cite{cond-exp1,cond-exp2,coldmol1,fesh-exp3,fesh-exp4} studies on 
reaction between ultracold molecules and atoms near Feshbach resonance. {\blue There have also been several studies on 
atom-molecule conversion by Feshbach resonance due to
coupled-channels \cite{rev}. 
However, the time-dependent sweeping across the Feshbach resonance to trigger conversion between atomic and molecular BEC has been rarely explored for systems involving multiple atomic modes.}


The goal of this paper is to provide a frame work to solve a dissociation mechanism 
in which different types of diatomic molecules undergo dissociation simultaneously, so that the individual reactions are coupled. 
The dynamics now involves multiple parameters and their interplay leads to various non-trivial effects such as interference and symmetry breaking, which may emerge due to competition between multiple modes corresponding to different types of atoms. Multiple modes can also describe higher degrees of freedom, i.e., spin angular momentum, rotational and vibrational modes.  

The general Hamiltonian describing reaction between atomic and molecular condensates  is given by
\begin{multline}
{\hat H}=\sum_{n} \epsilon_{n}^{\Psi} {\hat \Psi}_{n}^{\dagger}{\hat \Psi}_{n}+\sum_{k,n} \epsilon_{k,n}^{a} {\hat a}_{k,n}^{\dagger}{\hat a}_{k,n} +\epsilon_{k,n}^{b} {\hat b}_{k,n}^{\dagger}{\hat b}_{k,n} \\
+\sum_{k,n,m} g_{k,n,m}  {\hat \Psi}_n^{\dagger} {\hat a}_{k,m} {\hat b}_{k,n-m}+g_{k,n,m}^{*} {\hat \Psi}_n {\hat a}_{k,m}^{\dagger}{\hat b}_{k,n-m}^{\dagger},
    \label{H1}    
\end{multline}
where $k$ is the reaction channel~\cite{malla2022}, ${\hat \psi}({\hat \psi}^{\dagger})$, 
${\hat a}({\hat a}^{\dagger})$, and
${\hat b}({\hat b}^{\dagger})$ represent  annihilation  (creation) of the molecular field operators and atomic field operators, $n$ corresponding to a particular molecular mode. Hamiltonian (\ref{H1}), in general, is not analytically solvable. We propose to solve this model in the mean-field approximation where the molecular operators can be replaced by their expectation values \cite{yurovsky}. For this approximation to hold, we assume that the number of atoms produced is small compared to the number of molecules. We propose to solve these type of models in the Heisenberg picture, where the atomic field operators satisfy a {\it non-Hermitian} evolution equation.

There are two types of dissociation processes, stimulated dissociation and spontaneous dissociation. In the stimulated process an initially populated atomic mode (or modes) stimulates the dissociation of molecules. In the spontaneous process the atomic modes are initially empty and the molecules spontaneously dissociate into atoms. The spontaneous process could also describe cosmological production of particles and anti-particles due to vacuum fluctuations in the early stage of the Universe, which  may have been a consequence of fundamental parameters such as mass or interaction coefficients becoming time-dependent and the evolution passing through resonance \cite{fluctuation}.

After laying out the framework, we solve an illustrative example containing diatomic molecules with two atoms and each atom having two degrees of freedom; thus, the model includes four atomic modes. Let us denote the two degrees of freedom as spin-up and spin-down, however, note that they are bosons. We predict several key results for parameters satisfying the $CPT$ symmetry.  

First, the analytical result predicts an exponential sensitive imbalance of atomic population in each mode. For example, an atomic mode initially occupied with an spin-up atom stimulates an exponentially suppressed atomic production in spin-down states. This suppression is an artifact of destructive interference which arises due to the $CPT$ symmetry. However, a weakly broken symmetry alters the atomic production in each mode and can reverse the spin imbalance for a finite interference. 

Our second key result appears in the spontaneous process, where the chemical reactions are coupled, and the atoms are produced in a multi-mode squeezed state. Multi-mode squeezed states promise  applications in quantum metrology \cite{quantummetrology}, quantum communication, and quantum imaging \cite{quantumimag1,quantumimag2,quantumcomm}.  

In addition to these two central results, our solution  could open pathways for new investigations in time-dependent non-Hermitian systems.

\section{Theory of single dissociation process in the nonadiabatic regime} 
Let us start with the  dissociation process of a molecular BEC into two-mode atomic BEC, $AB\rightarrow A+B$. The dissociation process is described by the Hamiltonian in the curve crossing approximation,\cite{yurovsky,kayali}
\be
{\hat H}_{2}=\mu_1(t){\hat a}^{\dagger}{\hat a}+\mu_2(t){\hat b}^{\dagger}{\hat b}+J{\hat \psi}^{\dagger}{\hat a}{\hat b}+J^*{\hat \psi}{\hat a}^{\dagger}{\hat b}^{\dagger}.
\label{curve-1}
\ee
where the chemical potentials,  $\mu_i(t)$, of  two atomic modes are considered to time-dependent. In the nonadiabtic limit, only a small fraction of molecules get converted into atoms \cite{altland-LZ}. Assuming the number of molecules to be much larger than the number of atoms produced, we can replace the molecular field operator with the expectation value $\langle {\hat \psi}\rangle$\cite{yurovsky,kayali}. The system then reduces to interacting two atomic modes and the corresponding Hamiltonian reads
\be
{\hat H}_{eff}(t)=\mu_1(t){\hat a}^{\dagger}{\hat a}+\mu_2(t){\hat b}^{\dagger}{\hat b}+g{\hat a}{\hat b}+g^*{\hat a}^{\dagger}{\hat b}^{\dagger}.
\label{curve-2}
\ee
where $g=J\langle \psi^{\dagger} \rangle$.

The model (\ref{curve-2}) can be solved in the Heisenberg picture where the operators ${\hat a}(t)$ and ${\hat b}(t)$ satisfy   
\begin{align}
i \frac{d}{dt} \begin{pmatrix}
    {\hat a}(t)\\{\hat b}^{\dagger}(t)\end{pmatrix}=\begin{pmatrix} \mu_1(t) & g^* \\
-g & \mu_2(t)
\end{pmatrix}
 \begin{pmatrix}
    {\hat a}(t)\\{\hat b}^{\dagger}(t)\end{pmatrix}.
    \label{curve-3}
\end{align}
The molecular dissociation occurs near the crossing of two chemical potentials where the chemical potentials can be linearlized. Also, note that equation (\ref{curve-3}) is only applicable for bosonic atomic modes.

The number of atoms in $A$ mode at time $t\rightarrow \infty$ is then given by 
\begin{equation}
\langle {\hat a}^{\dagger}(t){\hat a}(t) \rangle_{t\rightarrow \infty} =n_{sp}^{(A)}+n_{st}^{(A)},
\end{equation}
where $n_{sp}$ corresponds to spontaneous dissociation while $n_{st}$ represents stimulated dissociation. For a linear sweep, $\mu_1(t)= \beta t,\mu_2(t)= -\beta t$, equation (\ref{curve-3}) resembles a non-Hermitian Landau-Zener (nLZ) model, and can be solved exactly  (see appendix A). The general solution of ${\hat a}(t)$ has the form ${\hat a}(t)=\phi_{A}(t){\hat a}(t_0)+\phi_{B}(t){\hat b}^{\dagger}(t_0)$, where $t_0$ represents the initial time and is assumed to be far away from the level crossing.  The asymptotic solution of $\phi_A(t\rightarrow \infty)$ and $\phi_B(t\rightarrow \infty)$,  with initial condition $\phi_{A}(t_0\rightarrow -\infty)=1$, $\phi_{B}(t_0\rightarrow -\infty)=0$, leads to the expressions
\begin{align}
|\phi_{A}(\infty)|^2=|\phi_{B}(\infty)|^2+1=\exp{\left(\frac{\pi|g|^2}{\beta}\right)}.
\label{nLZ}
\end{align}
In contrast to the Hermitian LZ model where $|\phi_{A}(\infty)|^2+|\phi_{B}(\infty)|^2=1$, the non-Hermiticity conserves the difference $|\phi_{A}(\infty)|^2-|\phi_{B}(\infty)|^2=1$.

The number of atoms in the spontaneous dissociation is
$n_{sp}^{(A)}=|\phi_{B}(\infty)|^2$. In the spontaneous dissociation process, unpairing is quantum correlated. The 
atoms are produced in a two-mode  squeezed state, where the maximum uncertainty along one direction is exponentially suppressed with the exponent proportional to  $|g|^2/|\beta|$.

The number of stimulated atoms in mode $A$ depends on the initial population of $A$ and $B$ modes and the phases of the solutions $\phi_{A}(t)$ and $\phi_{B}(t)$ at large times \cite{kayali}. If only the $A$ mode is initially populated with a single atom, then the number of atoms in the $A$ mode is $|\phi_{A}(\infty)|^2$ and in the $B$ mode is $|\phi_{B}(\infty)|^2$. Since the atoms are 
always produced in pairs the  population difference between $n_{st}^{(A)}$ and $n_{st}^{(B)}$ is conserved.


\begin{figure*}
\includegraphics[width=1\linewidth]{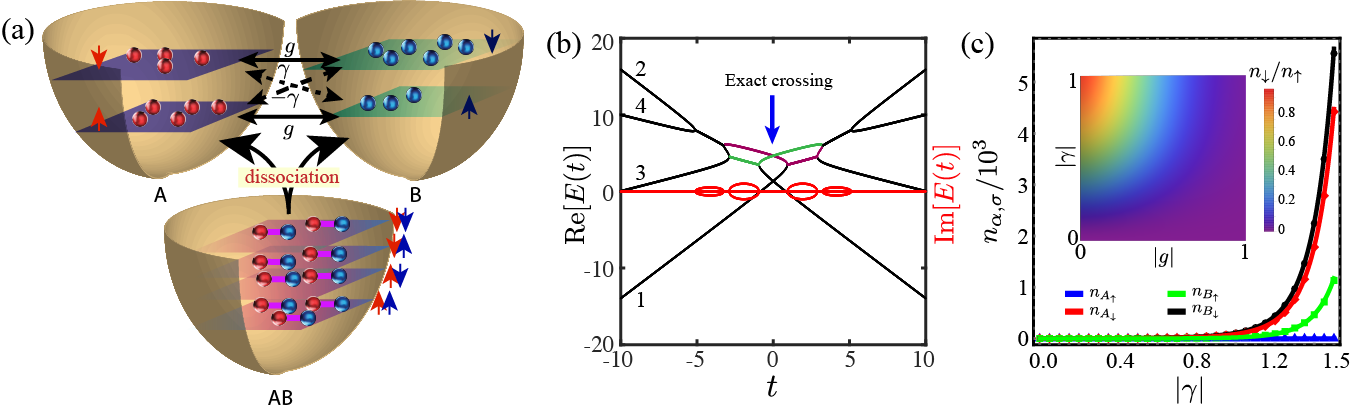}
\caption{(a) The schematic diagram of dissociation of diatomic molecules with four atomic modes.
(b) The instantaneous eigenvalues of (\ref{modelH}) are shown as a function of time for parameters $\beta_1=-1, \beta_2=0.5, E_1=5, E_2=1, g=0.5,\gamma=1$.  The black curves are real values and the red curves represent imaginary values. (c) The number of atoms produced in a particular mode when the system was stimulated by a single atom in $A_{\uparrow}$ mode. The scattered points are obtained from numerical evolution of (\ref{modelH}) while the solid curves are obtained from the analytical formulas in (\ref{unnormp}).}
\label{Fig1}
\end{figure*}

\section{Generalization to  dissociation of multiple molecular BECs} 
Now let us focus on a generic scenario where multiple molecular condensates undergo dissociation process. Multiple molecular condensates can consist of different types of molecules. These complex dynamics is difficult to handle. Here we propose a way to solve such a complex dynamics in the curve crossing approximation. 
The Hamiltonian describing dissociation of multiple molecular condensates consisting of diatomic molecules are given by
\begin{multline}
\label{Hgeneral}
{\hat H}_n(t)=\sum\limits_{i=1}^n \mu_{1}^{(i)}(t){\hat a}_i^{\dagger}{\hat a}_i+\sum\limits_{i=1}^m\mu_{2}^{(j)}(t){\hat b}_i^{\dagger}{\hat b}_ij
\\
+\sum\limits_{ij}\left[J_{ij}{\hat \psi}_{ij}^{\dagger}{\hat a}_i{\hat b}_j+h.c.\right],
\end{multline}
where  $\mu_1^{(i)}(t)$ and $\mu_2^{(i)}(t)$ are the time-dependent chemical potentials of  two-atomic modes for $i$-th type molecule, parameters $g_{i}$ are the coupling constants of the dissociation process $AB_{ij}\rightarrow A_{i}+B_{i}$.
The simultaneous dissociation of different types of molecules are correlated.

Hamiltonian (\ref{Hgeneral}) can describe dissociation of $n\times m$ types of different molecules. Employing similar procedures as before, if one replaces the molecular operators with their complex numbers then the system reduces many quadratically interacting atomic modes.  We can solve model (\ref{Hgeneral}) in the Heisenberg picture in a nonadiabatic regime. The solution requires solving an equation similar to (\ref{curve-3}), and the corresponding matrix satisfies non-Hermitian multi-state Landau-Zener (nMLZ). The Hermitian multi-state Landau-Zener (MLZ) models have been explored extensively, while the non-Hermitian counterpart has not been studied at all. In the next section we solve an illustrative example model of Hamiltonian (\ref{Hgeneral}) that includes four atomic modes.

\subsection{Example: dissociation process including four atomic modes} 
Now let us model a dissociation process, when different molecular condensates have same two atoms and each atom can be in different rotational or vibrational states. For simplicity, we restrict the atomic modes to two levels, and denote the levels as spin-up ($\uparrow$) and spin-down ($\downarrow$) states, see Fig. \ref{Fig1}a. Then the Hamiltonian describing the molecular dissociation process reads
\begin{align}
&{\hat H}_{4}= \mu_{1}(t) ({\hat a}_{\uparrow}^{\dagger}{\hat a}_{\uparrow}-{\hat a}_{\downarrow}^{\dagger}{\hat a}_{\downarrow})+\mu_{2}(t) ({\hat b}_{\uparrow}^{\dagger}{\hat b}_{\uparrow}-{\hat b}_{\downarrow}^{\dagger}{\hat b}_{\downarrow})
\nonumber\\
&+\Big[J_ 1{\hat \psi}_{\uparrow\uparrow}^{\dagger}{\hat a}_{\uparrow}{\hat b}_{\uparrow}+J_1{\hat \psi}_{\downarrow\downarrow}^{\dagger}{\hat a}_{\downarrow}{\hat b}_{\downarrow}
+J_2{\hat \psi}_{\uparrow\downarrow}^{\dagger}{\hat a}_{\uparrow}{\hat b}_{\downarrow}-J_2{\hat \psi}_{\downarrow\uparrow}^{\dagger}{\hat a}_{\downarrow}{\hat b}_{\uparrow}+h.c.\Big], 
    \label{curve-4}
\end{align}
where the  coupling parameters describing dissociation of triplet ($\uparrow\uparrow$) or ($\downarrow\downarrow$)  molecules are symmetric while the coupling parameters describing dissociation of singlet ($\uparrow\downarrow$) or ($\downarrow\uparrow$) molecules are anti-symmetric. This anti-symmetric nature is similar to a spin-orbit coupling \cite{NatureSO}. 

We solve (\ref{curve-4}) in the mean-field approximation and replace all the molecular operators with their expectation values. For a linear sweep, the evolution equation of the atomic field operators is given by $i \frac{d}{dt} {\hat \Phi}(t)=H_4^{(o)}(t) {\hat \Phi}(t)$, where the non-Hermitian matrix $H_4^{(o)}(t)$, in the  basis ${\hat \Phi}(t)=\{{\hat a}_{\uparrow}(t),{\hat a}_{\downarrow}(t), {\hat b}_{\uparrow}^{\dagger}(t),{\hat b}_{\downarrow}^{\dagger}(t)\}$, has  the form
\begin{equation}
H_4^{(o)}(t)
=\begin{pmatrix}
    b_1 t+E_1 & 0 & g^* & -\gamma^*\\0 & -b_1 t+E_1 & \gamma^* & g^*\\
    -g & -\gamma & b_2 t+E_2 & 0 \\\gamma & -g & 0 & -b_2t+E_2
    \end{pmatrix},
    \label{modelH}
\end{equation}
where the coupling parameters are  {\blue $g=J_1\langle{\hat \psi}_{\uparrow\uparrow}^{\dagger}\rangle=J_1\langle{\hat \psi}_{\downarrow\downarrow}^{\dagger}\rangle$ and $\gamma=-J_2\langle{\hat \psi}_{\uparrow\downarrow}^{\dagger}\rangle=-J_2\langle{\hat \psi}_{\downarrow\uparrow}^{\dagger}\rangle$, where $J_1$ and $J_2$ are real valued.} The parameters $b_1$ and $b_2$ are the slopes of the chemical potentials and $E_{i}$ correspond to level separations. Note that changing the slopes, level spacings, and coupling parameters one can explore various emergent mechanisms. 

\begin{figure*}
\includegraphics[width=0.9\linewidth]{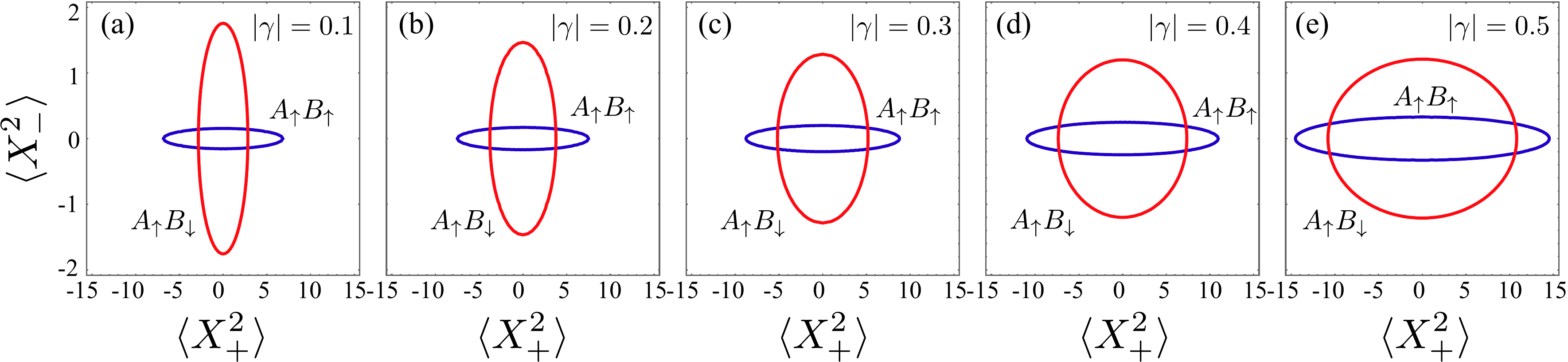}
\caption{The pairwise squeezing of $A$ and $B$ modes is shown from (\ref{SQ-1}) and (\ref{SQ-2}) and the constant shift is ignored in both directions. The parameters are $|g|=0.5$ and $|\beta_{1}|=|\beta_2|=1$. The expectation values are obtained from dimensionless operators ${\hat X}$.} 
\label{Sq}
\end{figure*}

The non-Hermitian model (\ref{modelH}) has a form similar to multi-state LZ (MLZ) model.  The
non-hermiticity leads to the emergence of complex eigenvalues near diabatic (diagonal) level crossings. The anti-Hermitian couplings ensure that the eigenvalues near level crossing are complex conjugate of each other. This regime is known as PT-symmetry broken phase in the non-Hermitian literature. Note that in the two-mode level crossing, there is one PT-broken phase \cite{Longstaff}. The model (\ref{modelH}), however, has four PT broken phases. In addition, the dynamics include two paths, purple and green colors in Fig. \ref{Fig1}b, that can amount to a finite path interference in the system. The interplay between the PT broken phase and the path interference strongly affects the final occupation of the atomic modes at large times. 

The solution of the evolution equation for  $H_{4}^{(o)}(t)$ requires finding the matrix $S$, 
which satisfies 
$
|\hat{\Phi}(t\rightarrow -\infty)\rangle=S |\hat{\Phi}(t\rightarrow +\infty)\rangle,
$
where $S\equiv U(T,-T)_{T\rightarrow \infty}$ is a non-unitary matrix. In Hermitian dynamics, $S$ is commonly known as the scattering matrix. The number of atoms in each mode depends on the matrix elements $S_{ij}$. These elements, however, cannot be obtained analytically for any arbitrary complex parameters $g$ and $\gamma$ \cite{Malla2017nonint,Malla2021,Bin2021}. 



{\blue In Ref.~\onlinecite{Nik2015} it was shown that the $CPT$ symmetry was partially responsible for the solvability of a Hermitian Hamiltonian similar to (\ref{modelH}) with real coupling parameters.  The three simultaneous operations in $CPT$ are defined as follows:\\
{\it (i)} time-reversal (${ T}$): change $t\rightarrow -t$, ${\hat a}\rightarrow {\hat a}^{\dagger}$, and 
${\hat b}^{\dagger}\rightarrow {\hat b}$;\\
{\it (ii)} complex conjugation ($ C$): change the imaginary number $i\rightarrow -i$, ${\hat a}\rightarrow {\hat a}^{\dagger}$, and 
${\hat b}^{\dagger}\rightarrow {\hat b}$;\\
{\it (iii)} parity (${ P}$): rename amplitudes ${\hat a}_{\uparrow}\rightarrow -{\hat a}_{\downarrow} $, ${\hat a}_{\downarrow}\rightarrow {\hat a}_{\uparrow}$, ${\hat b}_{\uparrow}\rightarrow -{\hat b}_{\downarrow} $, and ${\hat b}_{\downarrow}\rightarrow {\hat b}_{\uparrow}$. \\
It is straightforward to show that the evolution equation corresponding to the non-Hermitian model (\ref{modelH}) is invariant under $CPT$ transformation if the coupling parameters are real-valued, $g=g^*$ and $\gamma=\gamma^*$.}

The presence of  $CPT$ symmetry imposes the following relation between the elements of the $S$ matrix:
\be
{\hat S}={\hat S}'\equiv {\hat C}{\hat P} {\hat T} {\hat S}=\begin{pmatrix}
    S_{22} & -S_{12} & S_{42} & -S_{32}\\
    -S_{21} & S_{11} & -S_{41} & S_{31}\\
    S_{24} & -S_{14} & S_{44} & -S_{34}\\
    -S_{23} & S_{13} & -S_{43} & S_{33},
\end{pmatrix}
\label{scatter1}
\ee
where $S_{ij}$ are the matrix elements of ${\hat S}$ and the levels  \{1,2,3,4\} refer to the operators \{${\hat a}_{\uparrow},{\hat a}_{\downarrow},{\hat b}_{\uparrow}^{\dagger},{\hat b}_{\downarrow}^{\dagger}$\}, respectively. From (\ref{scatter1}) we obtain the following relations:
\begin{align}
S_{11}=S_{22}, S_{33}=S_{44},S_{12}=S_{21}=S_{34}=S_{43}=0.
    \label{relations}
\end{align}
The amplitude of the $S$ matrix elements $S_{ii}$, $S_{13}$, and $S_{24}$ can be obtained in the independent crossing approximation \cite{Dem}. The matrix elements $S_{14}$ and $S_{23}$ can be obtained under solvability, when one can substitute the level spacing parameters $E_{i}$ to be zero, and the model (\ref{modelH}) is equivalent to models with four modes crossing at a single point \cite{FLi2017}.

{\blue Our model has similarity with a particular subclass of  Hermitian multi-state Landau-Zener (MLZ) problems, and the properties of scattering matrices for the corresponding Hermitian model are listed in Ref.~\onlinecite{Nik2015}, where the parameters $g$ and $\gamma$ are considered real. In the Hermitian model, the transition probability is defined as $P_{ij}=|S_{ij}|^2$, and the probabilities satisfy the conservation of probability law, e.g., $\sum_{j} P_{ij}=1$. Here, $P_{ij}$ refers to the transition probability to find an electron in the $i$-th state at large times $t\rightarrow \infty$ starting from $j$-th state at initial time $t\rightarrow -\infty$.  The exact solution refers to the analytical formula equation (15) in Ref.~\onlinecite{Nik2015} for all the transition matrix elements.


Here, the non-Hermiticity destroys the unitarity and the 
transition probabilities do not add up to 1. Therefore, discussion of transition probability becomes irrelevant. 
Nevertheless, $|S_{ij}|^2$ still describes a physically relevant quantity and we define $n_{ij}=|S_{ij}|^2$, where $n_{ij}$ refers to the number of atoms produced in the $i$-th mode due to the dissociation process triggered by a single atom in the $j$-th mode. 

To obtain matrix elements $n_{ij}$, we apply similar principles of MLZ theory from Hermitian systems and treat the dynamics near each PT-broken phase as an individual nLZ transition. There are four individual PT-broken phases, with two of them controlled by $g$ while the other two are controlled by $\gamma$. Now we can replace the survival probability $p_g\equiv \exp(-\pi|g|^2/\beta_2)$ with ${\tilde n}_{g}\equiv \exp(\pi|g|^2/\beta_2)$ and the transition probability $q_g=1-p_g$ with ${\tilde n}_{g}-1$. With this substitution we find all the matrix elements of ${\hat n}$,
\begin{equation}
{\hat{n}}=
\begin{pmatrix}
    \tilde{n}_g \tilde{n}_{\gamma} & 0 & \tilde{n}_{\gamma}(\tilde{n}_g-1)& \tilde{n}_{\gamma}-1\\
    0 & \tilde{n}_g \tilde{n}_{\gamma} & \tilde{n}_{\gamma}-1 & \tilde{n}_{\gamma}(\tilde{n}_g-1) \\ \tilde{n}_{\gamma}(\tilde{n}_g-1) & \tilde{n}_{\gamma}-1 & \tilde{n}_g \tilde{n}_{\gamma} & 0 \\
    \tilde{n}_{\gamma}-1 & \tilde{n}_{\gamma}(\tilde{n}_g-1)& 0 & \tilde{n}_g \tilde{n}_{\gamma}
\end{pmatrix},
    \label{unnormp}
\end{equation}
where matrix elements in each column, with index $i$, represent the population in all the atomic levels when the reaction is triggered by a single atom in the $i$-th level. 
The matrix elements in a column satisfy the conservation law,
\begin{equation}
n_{jj}-\sum_{i}n_{ij}=1,
    \label{n-cons}
\end{equation}
where $j$-th mode is occupied with a single atom. This conservation law essentially implies that the atoms are produced in pairs and the population difference between $A$ and $B$ is an invariant. If the initial mode is occupied by  $n_0$ number of atoms then the number of atoms in each level is increased by $n_0$ times.
}



The number of stimulated atoms in each mode is shown in Fig. \ref{Fig1}c when $A_{\uparrow}$ is initially populated with a single atom. We find that our analytical prediction (\ref{unnormp}) perfectly agrees with the results obtained from numerical simulation of the evolution equation. The number of atoms in $A_{\downarrow}$ mode is zero, as predicted by the $CPT$ symmetry. This is an artifact of destructive interference due to the complete cancellation of two competing reaction channels, as shown in Fig. \ref{Fig1}b. The total number of down spins is purely produced in the $B_{\downarrow}$ mode. The ratio between the total number of down spin atoms and the total number of up spin atoms is then given by the  formula
\be
\frac{n_{\downarrow}}{n_{\uparrow}}=\frac{\tilde{n}_{\gamma}-1}{2\tilde{n}_g \tilde{n}_{\gamma}-\tilde{n}_{\gamma}}.
\label{ratio1}
\ee
For a small value of $|\gamma|^2/\beta_{1}$, the expression can be reduced to 
$
n_{\downarrow}/n_{\uparrow}=e^{-\frac{\pi |g|^2}{2|\beta_2|}} (|\gamma|^2/2|\beta_1|)$,
which is exponentially suppressed with parameter $|g|^2/|\beta_2|$ and depends linearly on parameter $|\gamma|^2/|\beta_1|$. This is one of our key result. This peculiar dependence is shown in the parametric plot in the inset of Fig. \ref{Fig1}c. 

{\blue
We also observe that the number of atoms in any mode only depends on $|S_{ij}|^2$.
Our model (\ref{modelH}) remains solvable even for the complex values of  $g$ and $\gamma$ when the phase difference between them is $\theta=n\pi$. In addition, the number of atoms in each model can be obtained from the same matrix ${\hat n}$ in (\ref{unnormp}), which was obtained using the $CPT$ symmetry for real values of $g$ and $\gamma$. This is possible since we can transform the evolution equation corresponding to the matrix  (\ref{modelH}) with complex $g$ and $\gamma$ to a new  evolution equation corresponding to a new matrix with real coupling parameters that satisfy $CPT$ symmetry when the phase difference between $g$ and $\gamma$ is $n\pi$. To understand this let us start with parameters of the form $g=|g|e^{i\phi}$ and $\gamma=|\gamma|e^{i\chi}$. Now the evolution equation reads
\begin{equation}
    i\frac{d {\hat \Phi}}{dt}=\begin{pmatrix}
    b_1 t+E_1 & 0 & |g|e^{-i\phi} & -|\gamma| e^{-i \chi}\\0 & -b_1 t+E_1 & |\gamma| e^{-i \chi} & |g|e^{-i\phi}\\
    -|g|e^{i\phi} & -|\gamma| e^{i \chi} & b_2 t+E_2 & 0 \\|\gamma| e^{i \chi} & -|g|e^{i\phi} & 0 & -b_2t+E_2
    \end{pmatrix}{\hat \Phi}.
    \label{complex-1}
\end{equation}
Next let us use the following transformation, ${\hat \Psi}=\{{\hat a}_{\uparrow}(t),{\hat a}_{\downarrow}(t), {\hat b}_{\uparrow}^{\dagger}(t)e^{-i\phi},{\hat b}_{\downarrow}^{\dagger}(t)e^{-i\phi}\}$. Then ${\hat \Psi}$ satisfies the equation of motion
\begin{equation}
    i\frac{d {\hat \Psi}}{dt}\hspace{-1mm}=\hspace{-1mm}\begin{pmatrix}
    b_1 t+E_1 & 0 & |g| & -|\gamma| e^{-i (\chi-\phi)}\\0 & -b_1 t+E_1 & |\gamma| e^{-i (\chi-\phi)} & |g|\\
    -|g| & -|\gamma| e^{i (\chi-\phi)} & b_2 t+E_2 & 0 \\|\gamma| e^{i (\chi-\phi)} & -|g| & 0 & -b_2t+E_2
    \end{pmatrix}\hspace{-1mm}{\hat \Psi}.
    \label{complex-2}
\end{equation}
Now if the phase difference $\chi-\phi=n\pi$, then the coupling parameters become real. This equation (\ref{complex-2}) is now invariant under $CPT$. Since the number of atoms corresponds to $\langle {\hat b}^{\dagger}{\hat b}\rangle$, the phase $e^{-i\phi}$ does not matter and the final atomic population can be given by (\ref{unnormp}).
}


\begin{figure*}
\includegraphics[width=1\linewidth]{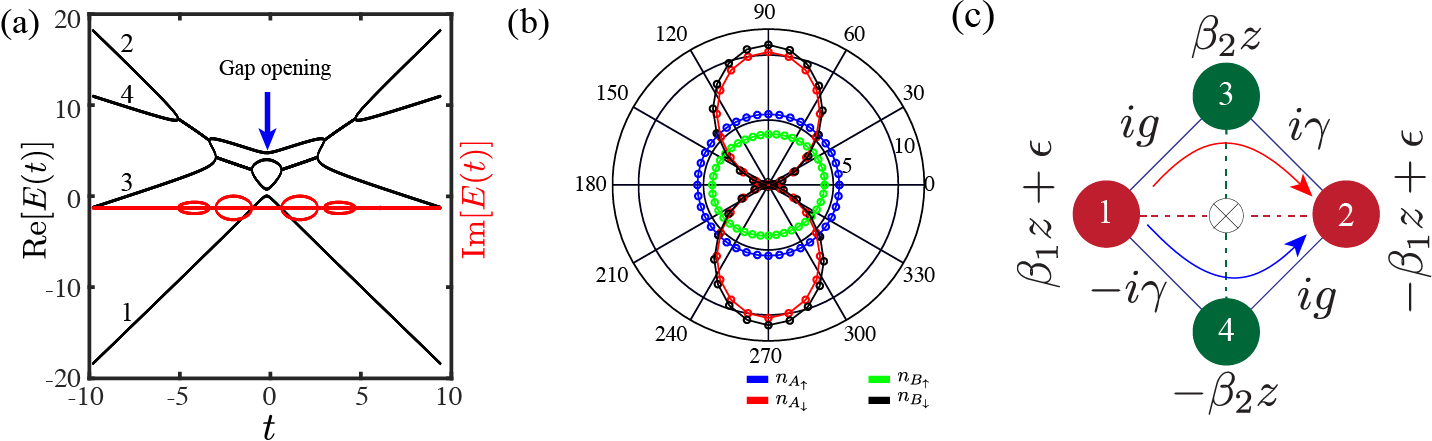}
\caption{(a) The instantaneous eigenvalues of (\ref{modelH}) are shown as a function of time for parameters $\beta_1=-1, \beta_2=0.5, E_1=5, E_2=1, g=0.5,\gamma=i$.  The black curves are real values and the red curves represent imaginary values. (b) The stimulated population of atomic modes from initially prepared single atom in $A_{\uparrow}$ is shown as a function of $\theta$. The colors blue, green, red, and black correspond to atomic modes $A_{\uparrow}$, $B_{\uparrow}$, $A_{\downarrow}$, and $B_{\downarrow}$, respectively. (c) The Schematic setup to realize the dynamics of our model  (\ref{modelH}). The red and green circles correspond to optical waveguides and the nearest coupling between waveguides are imaginary.}
\label{Fig3}
\end{figure*}

\subsection{Emergence of multi-mode squeezed states due to spontaneous emission} 
The spontaneous dissociation process in the four-mode model (\ref{modelH}) is quantum correlated. The number of atoms in each mode is equal due to the $CPT$ symmetry and can be obtained from 
(\ref{unnormp}),
\be
n_{\alpha,\sigma}={\tilde n}_g{\tilde n}_{\gamma}-1,
\label{spontaneous-four}
\ee
where $\alpha$ refers to $A$ or $B$ and $\sigma$ refers to the spin orientation. Another key quantum phenomenon emerging in the spontaneous dissociation process is that atoms are produced in a multi-mode squeezed state. The solvability of our model allows us to evaluate the amount of squeezing, and we calculate pair-wise squeezing of atomic modes. First, we observe that the modes $A_{\uparrow}$ and $A_{\downarrow}$ are not quantum correlated and are not in a two-mode squeezed state. The maximum uncertainty along both $X_{+}$ and $X_{-}$ is equal,  where $X$ is the position operator of the two-mode squeezed state. The same applies to modes $B_{\uparrow}$ and $B_{\downarrow}$.

Therefore, the atoms produced in the spontaneous process are always in a three-mode squeezed state \cite{MSQ}. To observe squeezing between $A_{\uparrow}$ and $B_{\uparrow}$ ($B_{\downarrow}$), we evaluate $\langle X_{\pm}^2\rangle$ (see appendix B) and find the maximum   values along each direction $X_{\pm}$ is given by
\begin{align}
  & \langle X_{\pm}^2\rangle_{A_{\uparrow} B_{\uparrow}}=c_{\gamma}+\frac{1}{2}|(\tilde{n}_{g}\tilde{n}_{\gamma})^{1/2}\pm ((\tilde{n}_{g}-1)\tilde{n}_{\gamma})^{1/2}|^2,~\hspace{-5mm}\label{SQ-1}\\
  & \langle X_{\pm}^2\rangle_{A_{\uparrow} B_{\downarrow}}=c_{g\gamma}+\frac{1}{2}|(\tilde{n}_{g}\tilde{n}_{\gamma})^{1/2}\pm (\tilde{n}_{\gamma}-1)^{1/2}|^2,~
\label{SQ-2}
\end{align}
where $c_{\gamma}=\frac{{\tilde n}_{\gamma}-1}{2}$ and $c_{g\gamma}=\frac{{\tilde n}_{\gamma}{\tilde n}_{g}-1}{2}$ are constant shifts.
In Fig. \ref{Sq} we plot $\langle X_{\pm}^2\rangle$ for different values of $|\gamma|$. The atoms produced in  $A_{\uparrow}$ and $B_{\uparrow}$ mode are always squeezed 
along the $X_{-}$ direction, and this phenomenon has the same origin as squeezing in a simple two-mode atomic dissociation. The atoms in $A_{\uparrow}$ and $B_{\downarrow}$ modes are squeezed along $X_{+}$ for small values of $|\gamma|$.  This squeezing is suppressed with an increase in $|\gamma|$ and eventually changes to  $X_{-}$ direction. {\blue Note that the direction of the squeezing is rotated by the phase $\phi$, since $\langle X^2 \rangle$ includes the correlator $\langle {\hat a} {\hat b} \rangle$.} From the symmetry of couplings, we find that the squeezing for up-spin $A$ and $B$ modes and down-spin $A$ and $B$ modes is equal, while the squeezing between up-spin $A$ and down-spin $B$ modes is the same as down-spin $A$ and up-spin $B$ modes.


\subsection{Phase difference between $g$ and $\gamma$ and broken solvability.--}
Now we discuss the scenario when there is a finite phase difference between  $g$ and $\gamma$. This finite phase difference, $\theta$, is responsible for violation of CPT symmetry, and gaps emerge near the exact crossing of levels with opposite slopes, see Fig. \ref{Fig3}a. This violation of exact crossing leads to asymmetric phase accumulations along the two interfering paths, leading to a finite interference in the system. Due to the finite interference, initial atoms in $A_{\uparrow}$ mode can stimulate the production of atoms in $A_{\downarrow}$ mode. However, an exact analytical solution is impossible, and we turn to a numerical approach. 

In Fig. \ref{Fig3}b, we show the $\theta$ dependence of the number of atoms in each mode stimulated by a single atom in $A_{\uparrow}$ mode. For $\theta=0$ or $\pi$ we recover the solvability condition and the $CPT$ symmetry of the evolution equation, which indeed agrees with formula (\ref{ratio1}), and the down-spin atoms are exponentially suppressed.
The production of down-spin atoms increases with increasing $\theta$, which corresponds to widening of the gaps and stronger constructive interference, see Fig. \ref{Fig3}b. Note, the imaginary parts of the eigenvalues, however, remain unaffected. At $\theta=(2m+1)\pi/2$, the interference becomes constructive and stimulates a maximum down spin. 

Concerning spontaneous dissociation, the total number atoms in each mode is no longer equal, however they still satisfy the condition $n_{A_{\uparrow}}+n_{A_{\downarrow}}=n_{B_{\uparrow}}+n_{B_{\downarrow}}$. The condition that atoms are produced in pairs ensures the quantum correlation between modes, and here the atoms will be produced in four-mode squeezing due to an effective coupling between $A_{\uparrow}$ and $A_{\downarrow}$.




\section{Discussion}
({\it i.}) We have solved a mechanism of dissociation of molecular condensates when there are competing chemical reactions due to multiple atomic modes. Our model can distinguish between the types of atoms produced in comparison to prior studies. {\blue The validity of our results relies on the fact that the molecular operators can be replaced with their expectation values. This assumption only describes a strong nonadiabatic conversion mechanism where only a small fraction of molecules can be converted to atoms. In our previous work Ref.~\onlinecite{malla2022} we analytically solved a model describing two-atomic modes where the exact solution allowed us to investigate both the nonadiabatic as well as the adiabatic regime. This work was motivated by the experiments in Ref.~\onlinecite{exp-21nat} where $Cs$ atoms were converted into $Cs_2$ molecules.

We only investigate the molecular dissociation process in the mean-field approximation and assume that molecules and atoms do not interact with each other. The experimental feasibility of such a system will depend on the preparation of molecular condensates in four modes. To prepare a simple multi-component condensate, one approach could be to prepare the molecules in different 
quantum states prior to condensation\cite{multiband}. This can be further simplified by assuming diatomic molecules consist of the same type of atoms, which will require us to prepare only 3-component condensates. When the molecules dissociate due to the time-varying chemical potentials, they will form atomic condensates. The number of atoms prepared is small compared to molecules are small and therefore the interaction between atoms can be neglected. A complete understanding requires understanding the stability of such a complex system which is beyond the scope of this article. 
}

({\it ii}) {\it Photonic realization of non-Hermitian multi-state Landau-Zener models}-
The similarity between the paraxial Helmoltz equation and the time-dependent Schr{\" o}dinger equation allows to implement various quantum mechanical models in photonic waveguides. To test various non-Hermitian dynamics, photonic waveguides provide a desirable platform due to the ability to control and manipulate the gain and loss of the media. The light propagation in a network of $N$ arbitrarily coupled waveguides satisfies coupled-mode equations in the form of the  paraxial Helmholtz equation. When the modes are orthogonal to each other, the paraxial Helmholtz equation resembles the time-dependent Schr{\"o}dinger equation 
$
i\partial \psi(z)/\partial z=H(z) \psi(z),
$
where the ``Hamiltonian" $H(z)$ is position dependent. To realize nMLZ models the refractive index in each waveguide can be made to be linearly dependent on the position ($z$) in order to design a particular setup of diabatic level crossings, similar to the one shown in Fig. \ref{Fig3}c. In order to introduce the nMLZ dynamics,  the coupling between the waveguides must be anti-Hermitian. The simplest way to satisfy anti-hermiticity is to consider the coupling to be imaginary. Recently, an imaginary coupling was realized in experiment~\cite{imag-coupling}, where the authors achieved the imaginary coupling between two waveguides by introducing an anchila waveguide in between them. This will allow for any possible arrangement to test a general nMLZ model. In Fig. \ref{Fig3}c, we show a design to test our model (\ref{modelH}). This design allows for imaginary coupling between two nearest neighbors and forbids coupling between diagonal waveguides. This design should be able to test the model (\ref{modelH}) for any set of parameters, and the interfering paths are shown by red and blue arrows. For our solvable model, one should expect the following observations. First, staring light at waveguide 1, one should observe zero intensity in the waveguide 2 at large distances. Second, in the large coupling limit one should observe equally strong intense light in 1 and 3. The intensity of light in the fourth waveguide will not be zero, however in comparison with 1 and 3, the intensity will be exponentially suppressed.

\section*{Acknowledgements}
R.K.M thanks Nikolai Sinitsyn, Avadh Saxena, Julia Cen, Mikhail Raikh, Wilton Kort-Kamp, and Saikat Banerjee for useful discussions. This work was  supported by the U.S. Department of Energy, Office of Science, Basic Energy Sciences, Materials Sciences and Engineering Division, Condensed Matter Theory Program, and the LANL  Center for Nonlinear Studies. 
\appendix
\section{Non-Hermitian Landau-Zener model} The non-Hermitian Landau-Zener (nLZ) model can be described by the matrix
\begin{equation}
{\hat H}_{nLZ}=\begin{pmatrix}
\beta t && g \\ -g^* && -\beta t
\end{pmatrix},
    \label{nLZ}
\end{equation}
where $\pm\beta$ are the two slopes corresponding to two levels and $g$ is the level coupling, which in general is complex. Unlike the Hermitian model the level couplings $g$ and $-g^*$ are negative complex conjugate of each other. The eigenvalues of ${\hat H}_{nLZ}$ are given by 
\begin{equation}
E_{nLZ}(t)=\pm \sqrt{\beta^2t^2-|g|^2}.
    \label{ELZ}
\end{equation}
In standard LZ model the eigenvalues are given by $E_{LZ}=\pm \sqrt{\beta^2t^2+|g|^2}$.
\begin{figure*}
\includegraphics[width=0.6\linewidth]{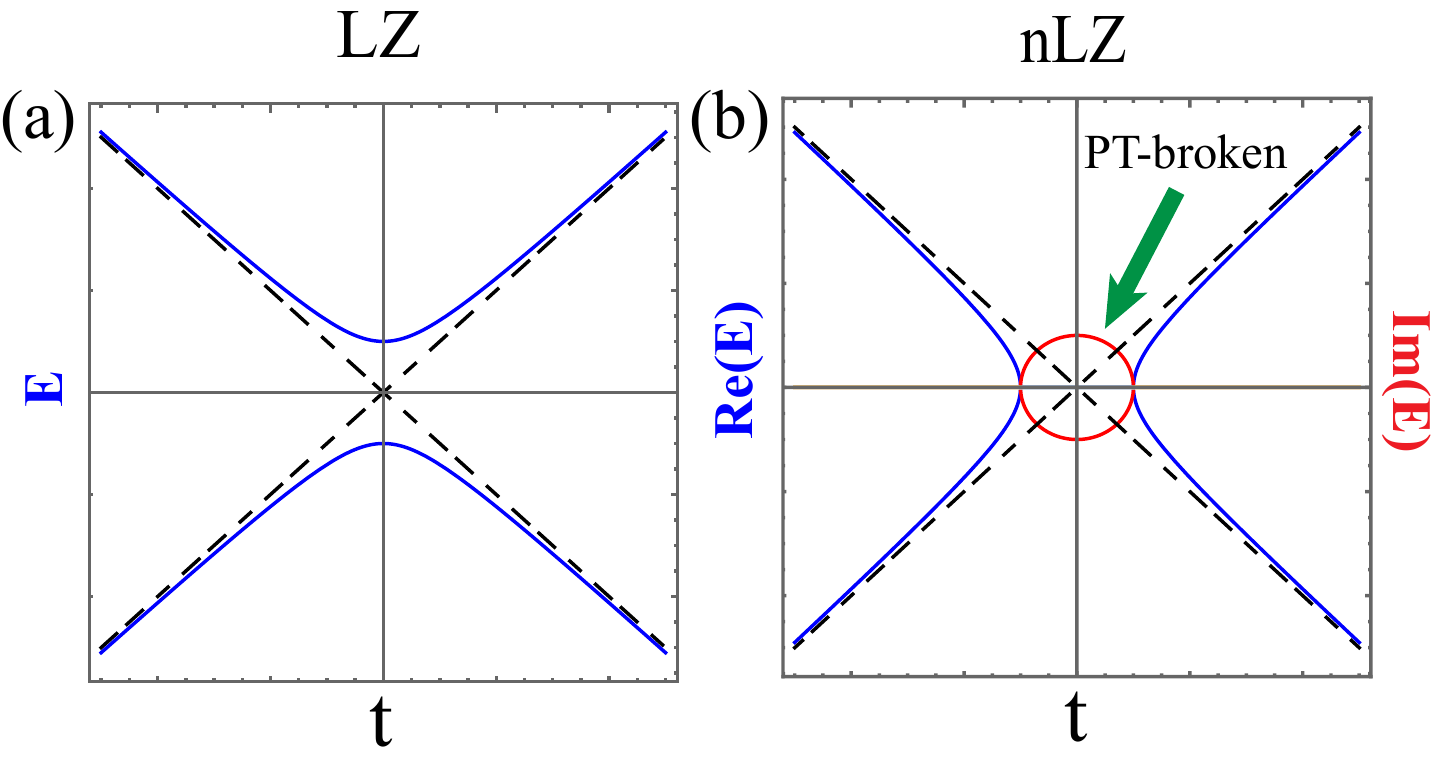}
\caption{The (a) eigen-energies of a standard LZ model and (b) the eigenvalues of a matrix describing the nLZ model ($E_{nLZ}(t)$) are schematically shown as a function of time. The blue corresponds to real part of eigenvalues while the red corresponds to the imaginary part. } 
\label{SM1}
\end{figure*}
The eigenvalues $E_{nLZ}(t)$ are shown as a function of time in Fig. \ref{SM1}b together with the eigen-energies for a standard LZ model with parameter $\beta$ and $g$.

The solution of the evolution equation for matrix (\ref{nLZ}) has the form of a $2\times 1$ column vector,
$$
|\phi(t)\rangle=\begin{pmatrix}
a(t) \\ b(t)
\end{pmatrix}, 
$$
where $a(t)$ satisfies a second order differential equation
\begin{equation}
\ddot{a}(t)+(\beta^2t^2-|g|^2+i\beta)a(t)=0,
    \label{second}
\end{equation}
whose solutions are given by parabolic cylinder functions \cite{parabolic}.

The solution can now be expressed as follows \cite{malla2017}
\begin{equation}
|\phi(t)\rangle=\phi_1\begin{pmatrix}
D_{\nu}(z) \\ -i\sqrt{\nu}D_{\nu-1}(z)
\end{pmatrix}+\phi_2\begin{pmatrix}
D_{\nu}(-z) \\ -i\sqrt{\nu}D_{\nu-1}(-z)
\end{pmatrix},
    \label{linear-comb}
\end{equation}
where $D_{\nu}(z)$ is the parabolic cylinder function, with $\nu=i|g|^2/2\beta$ and $z=\sqrt{2\beta}e^{i\pi/4}t$. Since we are only interested in the dynamics at large times $t\rightarrow \pm \infty$, the  asymptotic behavior of parabolic cylinder functions is good enough for our analysis.  
The critical difference between the Hermitian and the non-Hermitian dynamics comes from the phase of $\nu$, which is $-\pi/2$ for the Hermitian case and $\pi/2$ for the non-Hermitian case. This crucial difference leads to the following distinction between LZ and nLZ dynamics. 
Assuming the system starts with the upper state, $|a(t\rightarrow -\infty)|^2=1$, the asymptotic solution of $a(t)$ at large positive times is given by $|a(t\rightarrow \infty)|^2=e^{-\pi|g|^2/\beta}$ for the Hermitian model and $|a(t\rightarrow \infty)|^2=e^{\pi|g|^2/\beta}$ for the non-Hermitian model. Similarly, the solution $b(t)$ at large positive times is given by $|b(t\rightarrow \infty)|^2=1-e^{-\pi|g|^2/\beta}$ for the Hermitian model and $|b(t\rightarrow \infty)|^2=e^{\pi|g|^2/\beta}-1$ for the non-Hermitian model.

This exponential dependence in the non-Hermitian model appears due to the anti-Hermitian complex couplings. The conservation law in the Hermitian model describes the common probability law,
$$
|a(t)|^2+|b(t)|^2=1.
$$
The probability of survival is given by  $|a(t\rightarrow \infty)|^2$ while the probability of adiabatic transition is given by $|b(t\rightarrow \infty)|^2$. In the non-Hermitian model, however, there are no such terms, and the exponentially growing nature makes it difficult to call it probability. Nevertheless, in our model, $|a(t\rightarrow \infty)|^2$  and $|b(t\rightarrow \infty)|^2$ correspond to the number of atoms produced in the respective atomic modes. 
The conservation law here results from the asymptotic expansion of parabolic cylinder functions,
$$
|a(t)|^2-|b(t)|^2=1,
$$
and agrees with our dissociation mechanism in which atoms are produced in pairs.

\section{Pairwise squeezing of atomic modes}
The spontaneous dissociation process is quantum correlated and the atoms are produced in multi-mode quantum squeezed states. To quantify squeezing, we define position operators for the two modes corresponding to different types atoms $A$ and $B$. The position operator reads
\begin{equation}
{\hat X}_{A_{\sigma}B_{\sigma'}}^{(\phi)}=\frac{1}{2}\left[e^{-i\phi}({\hat a}_{\sigma}+{\hat b}_{\sigma'})+ e^{i\phi}({\hat a}_{\sigma}^{\dagger}+{\hat b}_{\sigma'}^{\dagger})\right],
    \label{O1}
\end{equation}
where $\sigma$, $\sigma'$ refer to the spins while $\phi$ is the angle that defines the measured quadrature. It is straightforward to see that the expectation value of operator $X$ is zero, while the expectation value of $X^2$ can be non-zero. We expand the expression for the $X^2$ operator from (\ref{O1})
\begin{multline}
{\hat X^{(\theta)^2}}_{\pm, A_{\sigma}B_{\sigma'}}=\frac{1}{4}\Big[e^{-2i\phi} ({\hat a}_{\sigma}^{2}+ {\hat b}_{\sigma'}^{2}+{\hat a}_{\sigma}{\hat b}_{\sigma'}+{\hat b}_{\sigma'}{\hat a}_{\sigma})+
\\
{\hat a^{\dagger}}_{\sigma}{\hat a}_{\sigma}+{\hat a^{\dagger}}_{\sigma}{\hat b}_{\sigma'}+
{\hat b^{\dagger}}_{\sigma'}{\hat a}_{\sigma}+{\hat b^{\dagger}}_{\sigma'}{\hat b}_{\sigma'}\Big]+h.c.
    \label{O2}
\end{multline}
At large times $t\rightarrow \infty$, the operators satisfy
\begin{align}
{\hat a}_{\uparrow}(t)=S_{11} {\hat a}_{\uparrow}(t_0) +S_{12} {\hat a}_{\downarrow}(t_0)+S_{13} {\hat b}_{\uparrow}^{\dagger}(t_0)+S_{14} {\hat b}_{\downarrow}^{\dagger}(t_0),\nonumber\\
{\hat a}_{\downarrow}(t)=S_{21} {\hat a}_{\uparrow}(t_0) +S_{22} {\hat a}_{\downarrow}(t_0)+S_{23} {\hat b}_{\uparrow}^{\dagger}(t_0)+S_{24} {\hat b}_{\downarrow}^{\dagger}(t_0),\nonumber\\
{\hat b}_{\uparrow}^{\dagger}(t)=S_{31} {\hat a}_{\uparrow}(t_0) +S_{32} {\hat a}_{\downarrow}(t_0)+S_{33} {\hat b}_{\uparrow}^{\dagger}(t_0)+S_{34} {\hat b}_{\downarrow}^{\dagger}(t_0),\nonumber\\
{\hat b}_{\downarrow}^{\dagger}(t)=S_{41} {\hat a}_{\uparrow}(t_0) +S_{42} {\hat a}_{\downarrow}(t_0)+S_{43} {\hat b}_{\uparrow}^{\dagger}(t_0)+S_{44} {\hat b}_{\downarrow}^{\dagger}(t_0)
    \label{O3}.
\end{align}
Here, $t_0$ corresponds to initial time. The non-zero contribution to $\langle X^2\rangle$ comes from terms proportional to $\langle {\hat a}(t_0){\hat a}^{\dagger}(t_0)\rangle$ and $\langle{\hat b}(t_0){\hat b}^{\dagger}(t_0)\rangle$. The term $\langle {\hat a}(t_0){\hat a}^{\dagger}(t_0)\rangle=1+\langle {\hat a}^{\dagger}(t_0) {\hat a}(t_0)\rangle$ and the second term vanishes since the initial atomic population is zero. 

Now we can express the expectation value of $X^2$ with only non-zero terms
\begin{equation}
\langle {\hat X^{(\theta)^2}}_{A_{\sigma}B_{\sigma'}} \rangle=\frac{1}{4}\left[e^{-2i\phi} ({\hat a}_{\sigma}{\hat b}_{\sigma'}+{\hat b}_{\sigma'}{\hat a}_{\sigma})+
{\hat a^{\dagger}}_{\sigma}{\hat a}_{\sigma}+{\hat b^{\dagger}}_{\sigma'}{\hat b}_{\sigma'}\right]+h.c.
    \label{O4}
\end{equation}

Let us first consider the expectation value of $X^2$ for $A_{\uparrow}$ and $B_{\uparrow}$ modes. Substituting equation (\ref{O3}) in equation (\ref{O2}) we find  
\begin{equation}
\langle {\hat X^{(\theta)^2}}_{\pm, A_{\uparrow}B_{\uparrow}} \rangle=\frac{1}{2}\left[Re[S_{11}S_{13}e^{2i\phi}]+ |S_{11}|^2+|S_{13}|^2+|S_{14}|^2\right],
    \label{AB-1}
\end{equation}
where
$$
-2|S_{11}||S_{13}| \leq Re[S_{11}S_{13}e^{2i\phi}]\leq 2|S_{11}||S_{13}|.
$$
 We can obtain the maximum and the minimum value of $\langle {\hat X^{(\theta)^2}}_{\pm, A_{\uparrow}B_{\uparrow}} \rangle$, which are given by
\begin{equation}
    \langle {\hat X^{(\theta)^2}}_{\pm, A_{\uparrow}B_{\uparrow}} \rangle=\frac{(\tilde{n}_{\gamma}-1)}{2}+\frac{1}{2}|(\tilde{n}_{g}\tilde{n}_{\gamma})^{1/2}\pm ((\tilde{n}_{g}-1)\tilde{n}_{\gamma})^{1/2}|^2,
\end{equation}
and we obtain our equation (20) from the main text. 
For $\gamma=0$ we recover the two-mode squeezing in the spin-independent atomic dissociation in Ref. \cite{yurovsky}. 

Similarly we can obtain equation (21) in our main text by switching $S_{13}$ and $S_{14}$ in equation (\ref{AB-1}). 

\bibliographystyle{apsrev4-1}
\bibliography{ref}
\vspace{12pt}

\end{document}